\documentclass[preprint,aps]{revtex4}

\usepackage{graphics}

\begin{document}

\title{WKB Analysis of the Scattering of Massive Dirac Fields in Schwarzschild Black Hole Spacetimes}

\author{H. T. Cho}
  \email{htcho@mail.tku.edu.tw}
\author{Y.-C. Lin}
\affiliation{Department of Physics, Tamkang University, Tamsui,
Taipei, Taiwan, Republic of China}

\date{\today}

\begin{abstract}
We analysis the radial equations for massive Dirac fields in
Schwarzschild black hole spacetimes. Different approximation
formulae under the WKB scheme are developed for the transmission
probability ${\cal T}$ of the radial wavefunction with $E^{2}\gg
V_{m}$, $E^{2}\approx V_{m}$, and $E^{2}\ll V_{m}$, where $E$ is
the energy of the field and $V_{m}$ is the maximum value of the
effective potential. Explicit results of ${\cal T}$ in these
approximations are given for various values of $E$, the mass $m$,
and the angular momentum parameter $\kappa$ of the fields. We also
discuss the dependence of ${\cal T}$ on these parameters.
\end{abstract}


\maketitle

\section{Introduction}

One of the most useful and efficient ways to study the properties
of black holes is by scattering matter waves off them
\cite{futterman}. From the more practical point of view, the study
of wave scattering in black hole spacetimes is crucial to the
understanding of the signals expected to be received by the new
generation of gravitational-wave detectors in the near future
\cite{hough}. Since the linear perturbations of black holes are
represented by fields of integral spins, the study of the
scattering of wave fields are concentrated on these cases while
that of the Dirac fields are thus less common, especially for the
massive ones \cite{futterman,unruh}. Recently, Finster and
collaborators \cite{finster} have renewed the interests in that of
the massive Dirac fields by investigating in details their
evolutions in various black hole spacetimes. Here we would like to
work in this direction by considering the scattering solutions of
the radial equations of massive Dirac fields in spherically
symmetric Schwarzschild black hole spacetimes.

There are both numerical and analytical methods in solving the
various wave equations in black hole scattering \cite{futterman}.
In this work we shall use the semi-analytic WKB approximation
\cite{berry,bender}, which has been proven to be very useful and
accurate in many cases like, for example, the evaluation of the
quasinormal mode frequencies \cite{schutz,iyer}. For the radial
Dirac equations we consider here, the effective potential can
change from a barrier to a step and vice versa when the mass $m$
or the angular momentum parameter $\kappa$ are varied. The WKB
scheme can be accommodated in various ways to consider all these
situations with different values of the energy $E$ of the field,
whether it is above or below the maximum value $V_{m}$ of the
potential.

In \cite{mukhopadhyay}, Mukhopadhyay and Chakrabarti have studied
in details the radial equations of a massive Dirac field in the
Schwarzschild black hole spacetime using a modified WKB (they
called it the "instantaneous WKB") approximation. However, they
only looked at the case with $m=M/2$, where $M$ is the mass of the
black hole, $\kappa=1$, and $E\approx m$. In this paper we use the
standard WKB scheme instead, and we shall extend the consideration
to cases with different values of $m$ and $\kappa$. In addition,
we shall consider all possible values of $E$ with $E\geq m$. In
this manner we can discuss the variations of the transmission
probabilities with respect to these various parameters.

In the next section, we consider the reduction of the massive
Dirac equation in Schwarzschild spacetimes into a set of
Schr\"odinger-like equations. We also discuss briefly the
properties of the corresponding effective potential \cite{cho}. In
Section III, we look at the three different WKB schemes for the
cases with $E^{2}\gg V_{m}$, $E^{2}\approx V_{m}$, and $E^{2}\ll
V_{m}$. Explicit approximated formulae for the transmission
probability are given. In Section IV, we apply the formulation to
calculate the transmission probabilities of the massive radial
Dirac equations for various values of $E$, $m$, and $\kappa$.
Conclusions and discussions are given in Section V.

\section{Dirac Equation in the Schwarzschild Spacetime}

In this section we discuss briefly the massive Dirac equation in
the Schwarzschild black hole spacetime, including its reduction to
a set of Schr\"odinger-like radial equations and the properties of
the corresponding effective potentials.

\subsection{Radial equations}

In the Schwarzschild spacetime,
\begin{equation}
ds^{2}=-\frac{\Delta}{r^2}dt^{2}+\frac{r^2}{\Delta}dr^{2} +r^{2}
d\theta^{2}+r^{2}{\rm sin}^{2}\theta\ d\phi^{2},
\end{equation}
where $\Delta=r(r-2M)$ and $M$ is the mass of the black hole.
Consider the Dirac equation in this background spacetime
\cite{brill},
\begin{equation}
[\gamma^{a}{e_{a}}^{\mu}(\partial_{\mu}+\Gamma_{\mu})+m]\Psi=0,
\label{diraceq}
\end{equation}
where $m$ is the mass of the Dirac field, and ${e_{a}}^{\mu}$ is
the inverse of the vierbein ${e_{\mu}}^{a}$ defined by the metric
$g_{\mu\nu}$,
\begin{equation}
g_{\mu\nu}=\eta_{ab}{e_{\mu}}^{a}{e_{\nu}}^{b},
\end{equation}
with $\eta_{ab}={\rm diag}(-1,1,1,1)$ being the Minkowski metric.
$\gamma^{a}$ are the Dirac matrices
\begin{equation}
\gamma^{0}= \left(
\begin{array}{cc}
-i&0\\0&i
\end{array}\right),\ \ \
\gamma^{i}=\left(
\begin{array}{cc}
0&-i\sigma^{i}\\i\sigma^{i}&0
\end{array}\right),\ i=1,2,3,
\end{equation}
where $\sigma^{i}$ are the Pauli matrices. $\Gamma_{\mu}$ is the
spin connection given by
\begin{equation}
\Gamma_{\mu}=\frac{1}{8}[\gamma^{a},\gamma^{b}]{e_{a}}^{\nu}e_{b\nu;\mu}\
,\label{spinconnection}
\end{equation}
where
$e_{b\nu;\mu}=\partial_{\mu}e_{b\nu}-\Gamma^{\alpha}_{\mu\nu}e_{b\alpha}$
is the covariant derivative of $e_{b\nu}$ with
$\Gamma^{\alpha}_{\mu\nu}$ being the Christoffel symbols.

Here it is convenient to choose the vierbein
\begin{equation}
{e_{\mu}}^{a} = \left(
\begin{array}{cccc}
\Delta^{1/2}/r&0&0&0\\ 0&r{\rm sin}\theta\ \!{\rm
cos}\phi/\Delta^{1/2} & r{\rm sin}\theta\ \!{\rm
sin}\phi/\Delta^{1/2} & r{\rm cos}\theta/\Delta^{1/2} \\ 0 & r{\rm
cos}\theta\ \!{\rm cos}\phi & r{\rm cos}\theta\ \!{\rm sin}\phi &
-r{\rm sin}\theta
\\ 0 & -r{\rm sin}\theta\ \!{\rm sin}\phi & r{\rm
sin}\theta\ \!{\rm cos}\phi & 0
\\
\end{array}\right).
\end{equation}
Since the spacetime is spherically symmetric, one can, after some
algebra, simplify the Dirac equation to \cite{brill,groves}
\begin{equation}
\frac{\gamma^{0}r}{\Delta^{1/2}}\frac{\partial\Psi}{\partial
t}+\frac{\tilde{\gamma}\Delta^{1/4}}{r^{3/2}}\frac{\partial}{\partial
r}(r^{1/2}\Delta^{1/4}\Psi)
-\frac{\tilde{\gamma}}{r}(\vec{\Sigma}\cdot\vec{L}+1)\Psi+m\Psi=0,
\end{equation}
where $\tilde{\gamma}$ is defined as
\begin{equation}
\tilde{\gamma}=\gamma^{1}{\rm sin}\theta\ \!{\rm cos}\phi\
\!+\gamma^{2}{\rm sin}\theta\ \!{\rm sin}\phi+\gamma^{3}{\rm
cos}\theta.
\end{equation}
Also
\begin{equation}
\vec{\Sigma}=\left(
\begin{array}{cc}
\vec{\sigma}&0\\0&\vec{\sigma}
\end{array}\right),
\end{equation}
and $\vec{L}=\vec{r}\times\vec{p}$ are the ordinary angular
momentum operators. The wavefunction $\Psi$ can be separated into
its radial and angular parts by writing
\begin{equation}
\Psi=\frac{1}{r^{1/2}\Delta^{1/4}}e^{-iEt}\Phi,
\end{equation}
where \cite{bjorken}
\begin{equation}
\Phi(r,\theta,\phi)= \left(\begin{array}{c}
\frac{iG^{(\pm)}(r)}{r}\varphi^{(\pm)}_{jm}(\theta,\phi)\\
\frac{F^{(\pm)}(r)}{r}\varphi^{(\mp)}_{jm}(\theta,\phi)\end{array}
\right),
\end{equation}
with the angular parts of the wavefunction
\begin{equation}
\varphi^{(+)}_{jm}=\left(
\begin{array}{c}
\sqrt{\frac{l+1/2+m}{2l+1}}{Y_{l}}^{m-1/2}\\
\sqrt{\frac{l+1/2-m}{2l+1}}{Y_{l}}^{m+1/2}
\end{array}\right),
\end{equation}
for $j=l+1/2$, and
\begin{equation}
\varphi^{(-)}_{jm}=\left(
\begin{array}{c}
\sqrt{\frac{l+1/2-m}{2l+1}}{Y_{l}}^{m-1/2}\\
-\sqrt{\frac{l+1/2+m}{2l+1}}{Y_{l}}^{m+1/2}
\end{array}\right),
\end{equation}
for $j=l-1/2$.

Then the radial equations \cite{chandrasekhar,cho} can be written
as
\begin{eqnarray}
\left(-\frac{d^{2}}{dx^{2}}+V_{1}\right)\hat{F}
&=&E^{2}\hat{F},\label{V1}\\
\left(-\frac{d^{2}}{dx^{2}}+V_{2}\right)\hat{G}
&=&E^{2}\hat{G},\label{V2}
\end{eqnarray}
where
\begin{equation}
V_{1,2}=\pm\frac{dW}{dx}+W^{2},\label{V12}
\end{equation}
with
\begin{equation}
W=\frac{\Delta^{1/2}(\kappa^{2}+m^{2}r^{2})^{3/2}}{r^{2}(\kappa^{2}+m^{2}r^{2})+m\kappa
\Delta/2E}, \label{W}
\end{equation}
where
\begin{equation}
x=r+2M{\rm ln}\left(\frac{r}{2M}-1\right)+\frac{1}{2E}{\rm
tan}^{-1}\left(\frac{mr}{\kappa}\right).
\end{equation}
Here we have combined the $(+)$ and the $(-)$ cases, with $\kappa$
going over all positive and negative integers. Positive integers
represent the $(+)$ cases with
\begin{equation}
\kappa=j+\frac{1}{2}\ \ \ {\rm and}\ \ \  j=l+\frac{1}{2},
\end{equation}
and
\begin{equation}
\left(
\begin{array}{c}
\hat{F}\\ \hat{G} \end{array} \right) =\left(
\begin{array}{cc}
{\rm sin}(\theta/2)&{\rm cos}(\theta/2)\\ {\rm
cos}(\theta/2)&-{\rm sin}(\theta/2)
\end{array}\right)
\left(
\begin{array}{c}
F\\G
\end{array}
\right),
\end{equation}
where
\begin{equation}
\theta={\rm tan}^{-1}(mr/\kappa).
\end{equation}
While negative integers represent the $(-)$ cases with
\begin{equation}
\kappa=-\left(j+\frac{1}{2}\right)\ \ \ {\rm and}\ \ \
j=l-\frac{1}{2},
\end{equation}
and
\begin{equation}
\left(
\begin{array}{c}
\hat{F}\\ \hat{G} \end{array} \right) =\left(
\begin{array}{cc}
{\rm cos}(\theta/2)&-{\rm sin}(\theta/2)\\ {\rm
sin}(\theta/2)&{\rm cos}(\theta/2)
\end{array}\right)
\left(
\begin{array}{c}
F\\G
\end{array}
\right).
\end{equation}

From the Schr\"odinger-like equations in Eqs.~(\ref{V1}) and
(\ref{V2}), we shall consider the scattering of the massive Dirac
fields. Note that $V_{1}$ and $V_{2}$, which are related as shown
in Eq.~(\ref{V12}), are supersymmetric partners derived from the
same superpotential $W$ \cite{cooper}. It has been shown that
potentials related in this way possess the same spectra, discrete
as well as continuous \cite{anderson}. Physically this just
indicates that Dirac particles and antiparticles will scatter in
the same manner around the Schwarzshild black hole. We shall
therefore concentrate just on Eq.~(\ref{V1}) with potential
$V_{1}$ in the following sections.

\subsection{Properties of the effective potential}

Here we discuss briefly the dependence of the effective potential
\begin{eqnarray}
&&V(r,\kappa,m,E)\nonumber\\
&=&\frac{\Delta^{1/2}(\kappa^{2}+m^{2}r^{2})^{3/2}}
{(r^{2}(\kappa^{2}+m^{2}r^{2})+m\kappa\Delta/2E)^{2}}
\left[\Delta^{1/2}(\kappa^{2}+m^{2}r^{2})^{3/2}+((r-1)(\kappa^{2}+m^{2}r^{2})+
3m^{2}r\Delta)\right]\nonumber\\ &&\ \ \
-\frac{\Delta^{3/2}(\kappa^{2}+m^{2}r^{2})^{5/2}}{(r^2(\kappa^{2}+m^{2}r^{2})+m\kappa\Delta/2E)^{3}}
\left[2r(\kappa^{2}+m^{2}r^{2})+2m^{2}r^{3}+m\kappa(r-1)/E\right].
\label{massiveV}
\end{eqnarray}
on the parameters $m$, $\kappa$, and $E$. Since we shall only work
with $V_{1}$, but not $V_{2}$, we have dropped the subscript of
$V$.

\begin{figure}[!]
\includegraphics{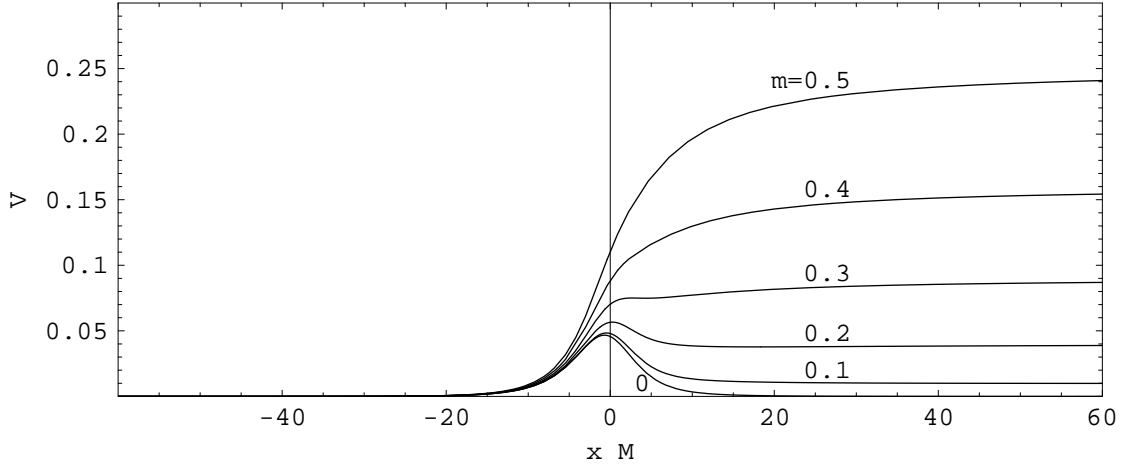}
\caption{\label{potentialwithm} Variation of the effective
potential with the mass $m$ (in units of $M$) of the Dirac field
for $\kappa=1$ and $E=M$.}
\end{figure}

First, its dependence on $m$ is showed in
Fig.~\ref{potentialwithm} with energy $E=M$ and with $\kappa=1$.
Note that we have shown $V$ as a function of $x$ in the figure.
For small values of $m$, the potential is in the form of a
barrier, with the asymptotic value
\begin{equation}
V(x\rightarrow\infty)=m^2.
\end{equation}
As $m$ is increased, the peak of the potential also increases but
does so very slowly. Eventually, the height of the peak is lower
than the asymptotic value $m^2$. When $m$ is increased further,
the peak disappears altogether, and the potential barrier turns
effectively into a potential step.

The effective potential $V$ also depends on the energy $E$.
However, as shown in \cite{cho}, the general behaviors of the
potential remain the same as $E$ is increased from its minimum
value $m$. Indeed, from the form of the potential in
Eq.~(\ref{massiveV}), $E$ appears all in the denominators. The
terms involving $E$ can never get large enough to change the
general behaviors of the potential since $E$ cannot be smaller
than $m$.

\begin{figure}[!]
\includegraphics{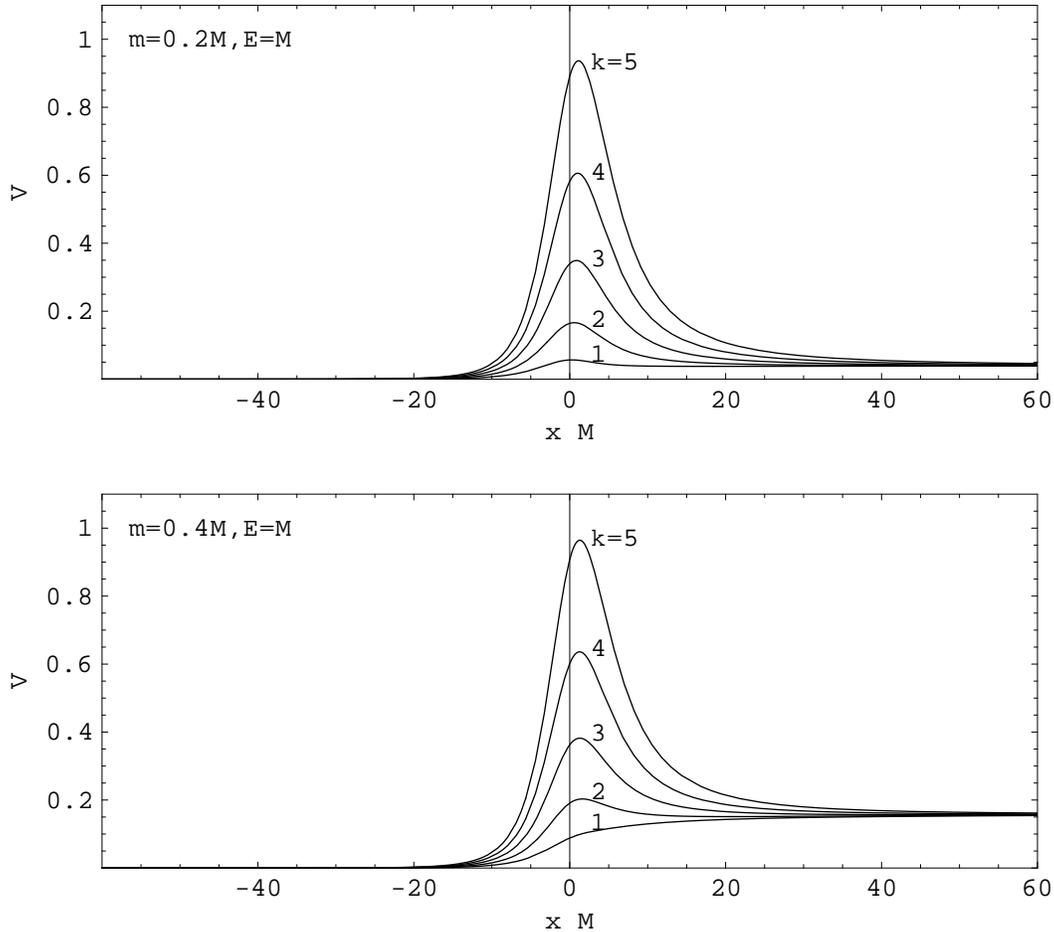}
\caption{\label{potentialwithk} Variation of the effective
potential with $\kappa$ for the Dirac fields of $m=0.2M$ and
$0.4M$.}
\end{figure}

Finally, in Fig.~\ref{potentialwithk} we show the dependence of
the effective potential with the angular momentum quantum number
$\kappa$. We see that as $\kappa$ increases, the behaviors of the
potential approach to that of the massless one, regardless of the
mass of the field. This can be seen by taking the limit
$|\kappa|\rightarrow\infty$ in Eq.~(\ref{massiveV}),
\begin{equation}
V(|\kappa|\rightarrow\infty)\approx\frac{\Delta\kappa^{2}}{r^{4}},
\end{equation}
which is independent of $m$.

\section{WKB approximations}

In this section we consider the radial equations obtained in the
previous section as a general one-dimensional quantum mechanical
problem,
\begin{equation}
\left(-\frac{d^{2}}{dx^{2}}+V\right)\psi
=E^{2}\psi,\label{schrodinger}
\end{equation}
with the asymptotic values of the potential
\begin{equation}
V(x\rightarrow\infty)=m^{2}\ \ \ \ \ {\rm and}\ \ \ \ \
V(x\rightarrow -\infty)=0.
\end{equation}
Suppose an incident wave comes from the right ($x=\infty$). Then
the boundary conditions for $\psi$ are
\begin{eqnarray}
\psi(x\rightarrow\infty)&=&e^{-i\sqrt{E^{2}-m^{2}}\ \!
x}+Re^{i\sqrt{E^{2}-m^{2}}\ \!x}\label{bcinf}\\ \psi(x\rightarrow
-\infty)&=&Te^{-iEx},\label{bcneginf}
\end{eqnarray}
where $R$ and $T$ represent the reflection and transmission
coefficients, respectively.

Since exact solutions for Eq.~(\ref{schrodinger}) are usually hard
to find, we have to resort to approximations. Here we adopt the
WKB approximation, which has been proved to be extremely useful
and sometimes to be more accurate than expected. We try to develop
approximate expressions for the transmission probability
\begin{equation}
{\cal T}=\sqrt{\frac{E^{2}}{E^{2}-m^{2}}}\
|T|^{2},\label{transdef}
\end{equation}
for the whole range of the energy $E$, including the cases with
$E^{2}\gg V_{m}$, $E^{2}\approx V_{m}$, and $E^{2}\ll V_{m}$,
where $V_{m}$ is the maximum value of the potential. In the case
of a barrier, $V_{m}$ will be the peak value of $V$, while in the
case of a step, $V_{m}=m^{2}$.

\subsection{$E^{2}\gg V_{m}$}

For $E^{2}\gg V_{m}$, the standard WKB form of the wavefunction
can be given by
\begin{equation}
\psi(x)=C_{+}W_{+}(x)+C_{-}W_{-}(x),\label{wkb}
\end{equation}
for $-\infty<x<\infty$, with
\begin{equation}
W_{\pm}(x)=\frac{1}{\sqrt{p(x)}}e^{\pm i\int^{x}_{x_{0}}dx'p(x')}
\end{equation}
being the WKB wavefunctions. Here $p(x)=\sqrt{E^{2}-V(x)}$,
$x_{0}$ is some fixed reference point, and $C_{+}$ and $C_{-}$ are
constants to be determined from the boundary conditions.
Eq.~(\ref{wkb}) is a useful approximation for the wavefunction as
long as the validity condition
\begin{equation}
\left\vert\frac{V'}{2p^{3}}\right\vert\ll 1,\label{validity}
\end{equation}
where $V'=dV/dx$, is fulfilled.

From the boundary condition at $x\rightarrow -\infty$
(Eq.~(\ref{bcneginf})), we have $C_{+}=0$, while from the
condition at $x\rightarrow\infty$ (Eq.~(\ref{bcinf})), we have
\begin{equation}
C_{-}=(E^{2}-m^{2})^{1/4}e^{-i\sqrt{E^{2}-m^{2}}\ \!x_{0}}
e^{i\int^{\infty}_{x_{0}}dx'(p(x')-\sqrt{E^{2}-m^{2}})}.
\end{equation}
Therefore, the transmission coefficient
\begin{eqnarray}
T&=&\frac{C_{-}}{\sqrt{E}}e^{iEx_{0}}e^{i\int^{x_{0}}_{-\infty}dx'(p(x')-E)}\nonumber\\
&=&\left(\frac{E^{2}-m^{2}}{E^{2}}\right)^{1/4}e^{i(E-\sqrt{E^{2}-m^{2}})x_{0}}
e^{i\int^{\infty}_{x_{0}}dx'(p(x')-\sqrt{E^{2}-m^{2}})}e^{i\int^{x_{0}}_{-\infty}dx'
(p(x')-E)},
\end{eqnarray}
and the transmission probability in Eq.~(\ref{transdef}) becomes
\begin{equation}
{\cal T}=1.
\end{equation}
This is of course consistent with the classical result in which
the particle moves to the left without rebounding. However,
quantum mechanically ${\cal T}$ is not exactly equal to 1, there
is always a small probability for reflection. Thus to get a better
approximation on ${\cal T}$ in this quantum mechanical situation,
we need to extend the WKB scheme used here.

Suppose we write the general solution of Eq.~(\ref{schrodinger})
as \cite{berry}
\begin{equation}
\psi=C_{+}(x)W_{+}(x)+C_{-}(x)W_{-}(x),\label{general}
\end{equation}
where $C_{+}(x)$ and $C_{-}(x)$ are now functions of $x$. If we
also take
\begin{equation}
C_{+}'(x)W_{+}(x)+C_{-}'(x)W_{-}(x)=0,
\end{equation}
and together with Eq.~(\ref{general}), we can obtain the
differential equations for $C_{\pm}$,
\begin{equation}
C_{\pm}'(x)=\mp
i\left(\frac{p''(x)}{4p(x)^2}-\frac{3(p'(x))^{2}}{8p(x)^{3}}\right)\left[C_{\pm}(x)+C_{\mp}(x)e^{\mp
2i\int^{x}_{x_{0}}dx' p(x')}\right]\label{deforc}
\end{equation}
From the boundary conditions in Eqs.~(\ref{bcinf}) and
(\ref{bcneginf}), we have
\begin{eqnarray}
C_{+}(-\infty)&=&0,\\
C_{-}(\infty)&=&(E^{2}-m^{2})^{1/4}e^{-i\sqrt{E^{2}-m^{2}}\
\!x_{0}}e^{i\int^{\infty}_{x_{0}}dx'
(\sqrt{E^{2}-V(x')}-\sqrt{E^{2}-m^{2}})}.
\end{eqnarray}
Finally, with these boundary values, the differential equations in
Eq.~(\ref{deforc}) can be turned into integral equations,
\begin{eqnarray}
C_{+}(x)&=&-i\int^{x}_{-\infty}dx'
\left(\frac{p''(x')}{4p(x')^2}-\frac{3(p'(x'))^{2}}{8p(x')^{3}}\right)\left[C_{+}(x')+C_{-}(x')e^{-
2i\int^{x'}_{x_{0}}dx'' p(x'')}\right],\\
C_{-}(x)&=&C_{-}(\infty)-
i\int^{\infty}_{x'}dx'\left(\frac{p''(x')}{4p(x')^2}-\frac{3(p'(x'))^{2}}{8p(x')^{3}}\right)
\left[C_{-}(x')+C_{+}(x')e^{ 2i\int^{x'}_{x_{0}}dx''
p(x'')}\right].\nonumber\\
\end{eqnarray}
To the lowest order, we just substitute the boundary values of
$C_{\pm}$ into the right hand side of the above equations,
\begin{eqnarray}
C_{+}(x)&\approx&-iC_{-}(\infty)\int^{x}_{-\infty}dx'
\left(\frac{p''(x')}{4p(x')^2}-\frac{3(p'(x'))^{2}}{8p(x')^{3}}\right)e^{-
2i\int^{x'}_{x_{0}}dx'' p(x'')},\label{cplus}\\
C_{-}(x)&\approx&C_{-}(\infty)\left[1-
i\int^{\infty}_{x'}dx'\left(\frac{p''(x')}{4p(x')^2}
-\frac{3(p'(x'))^{2}}{8p(x')^{3}}\right)\right].\label{cminus}
\end{eqnarray}
Under this approximation, the reflection coefficient becomes
\begin{equation}
R=\frac{C_{+}(\infty)}{(E^{2}-m^{2})^{1/4}}e^{-i\sqrt{E^{2}-m^{2}}\
\!x_{0}}e^{i\int^{\infty}_{x_{0}}dx'(\sqrt{E^{2}-V(x')}-\sqrt{E^{2}-m^{2}})},
\end{equation}
with
\begin{equation}
C_{+}(\infty)=-iC_{-}(\infty)\int^{\infty}_{-\infty}dx'
\left(\frac{p''(x')}{4p(x')^2}-\frac{3(p'(x'))^{2}}{8p(x')^{3}}\right)e^{-
2i\int^{x'}_{x_{0}}dx'' p(x'')},
\end{equation}
as given in Eq.~(\ref{cplus}). The reflection probability in this
approximation becomes
\begin{eqnarray}
{\cal R}&=&|R|^{2}\nonumber\\
&=&\left\vert\int^{\infty}_{-\infty}dx'
\left(\frac{p''(x')}{4p(x')^2}-\frac{3(p'(x'))^{2}}{8p(x')^{3}}\right)e^{-
2i\int^{x'}_{x_{0}}dx'' p(x'')}\right\vert^{2},
\end{eqnarray}
while the transmission coefficient is given by
\begin{eqnarray}
{\cal T}&=&1-{\cal R}\nonumber\\
&=&1-\left\vert\int^{\infty}_{-\infty}dx'
\left(\frac{p''(x')}{4p(x')^2}-\frac{3(p'(x'))^{2}}{8p(x')^{3}}\right)e^{-
2i\int^{x'}_{x_{0}}dx'' p(x'')}\right\vert^{2}.\label{e2ggvm}
\end{eqnarray}
This constitutes our WKB approximation for $E^{2}\gg V_{m}$ as
long as the validity condition in Eq.~(\ref{validity}) is
satisfied. Note that we can also obtain, in this approximation,
the wavefunction by substituting $C_{+}(x)$ and $C_{-}(x)$ in
Eqs.~(\ref{cplus}) and (\ref{cminus}), respectively, into
Eq.~(\ref{general}).

\subsection{$E^{2}\approx V_{m}$}

In the case of a potential barrier, when the energy $E$ of the
field is decreased to such an extend that $E^{2}$ is close to the
peak of the potential, the validity condition
(Eq.~(\ref{validity})) will no longer be satisfied near $x_{m}$,
the position of the peak. This situation can be remedied by
representing the part of the potential near $x_{m}$ as a parabola,
while maintaining the WKB solutions on either side of it. Exact
solutions can be found for the parabolic potential, and then the
approximate solution for the whole range, $-\infty<x<\infty$, can
be obtained by matching the three solutions across the
intertwining regions \cite{berry,bender}.

For the part of the potential near $x_{m}$, we can write, in the
parabolic approximation,
\begin{equation}
V(x)\approx V_{m}+\frac{1}{2}V''(x_{m})(x-x_{m})^{2},
\end{equation}
and
\begin{eqnarray}
E^{2}-V(x)&\approx& (E^{2}-V_{m})+\lambda(x-x_{m})^{2}\nonumber\\
&\approx& \left\{
\begin{array}{c}
\sqrt{\lambda}(z^{2}+\xi^{2})\ \ \ \ \ E^{2}>V_{m}\\
\sqrt{\lambda}(z^{2}-\xi^{2})\ \ \ \ \ E^{2}<V_{m}
\end{array}\right.
\end{eqnarray}
where
\begin{equation}
\lambda=-V''(x_{m})/2,\ \ \ \ \ z=\lambda^{1/4}(x-x_{m}),\ \ \ \ \
{\rm and} \ \ \ \ \ \xi=\frac{|E^{2}-V_{m}|^{1/2}}{\lambda^{1/4}}.
\end{equation}
Note that for
\begin{equation}
z=\pm\xi\Rightarrow
x_{1,2}=x_{m}\pm\frac{|E^{2}-V_{m}|^{1/2}}{\sqrt{\lambda}}.
\end{equation}
When $E^2$ is smaller than $V_{m}$, $x_{1}$ and $x_{2}$ are just
the turning points with $E^2=V(x_{1,2})$. In any case for
$E^{2}\approx V_{m}$, we can divide $x$ into three regions: (I)
$x>x_{1}$, (II) $x_{1}>x>x_{2}$, and (III) $x<x_{2}$. For regions
(I) and (III), we still use the WKB form of the wavefunction,
\begin{eqnarray}
\psi_{I}(x)&=&A_{+}W_{+}(x) +A_{-}W_{-}(x),\nonumber\\
\psi_{III}(x)&=&BW_{+}(x).
\end{eqnarray}
For region II, we have, in the parabolic approximation, the
Schr\"odinger equation in Eq.~(\ref{general}) can be written as
\begin{equation}
\frac{d^{2}\psi}{dz^{2}}+(z^{2}\pm\xi^{2})\psi=0,
\end{equation}
with the general solution
\begin{equation}
\psi_{II}(z)=\alpha
D_{-\frac{1}{2}\mp\frac{i\xi^{2}}{2}}(\sqrt{2}e^{i\pi/4}z)+\beta
D_{-\frac{1}{2}\mp\frac{i\xi^{2}}{2}}(-\sqrt{2}e^{i\pi/4}z),
\end{equation}
where $D_{\nu}(t)$ are the parabolic cylinder functions.

Taking into account of the boundary conditions in
Eqs.~(\ref{bcinf}) and (\ref{bcneginf}), and matching the
wavefunctions in different regions across $x=x_{1}$ and $x_{2}$,
we can solve for the constants $A_{+}$, $A_{-}$, $\alpha$,
$\beta$, and $B$ \cite{bender,iyer}. One can then obtain
\begin{eqnarray}
{\cal T}=\left\{
\begin{array}{l}
1/(1+e^{-\pi\xi^{2}})\ \ \ \ \ E^{2}>V_{m}\\ 1/(1+e^{\pi\xi^{2}})\
\ \ \ \ \ E^{2}<V_{m}
\end{array}\right.
\end{eqnarray}
These two cases can be combined by writing
\begin{equation}
{\cal
T}=\frac{1}{1+e^{\pi(V_{m}-E^{2})/\sqrt{\lambda}}},\label{e2appvm}
\end{equation}
which is the WKB approximation we shall use for $E^{2}\approx
V_{m}$.

\subsection{$E^{2}\ll V_{m}$}

When the energy $E$ of the field is lowered further, the turning
points $x_{1}$ and $x_{2}$ will move far apart in such a way that
the parabolic approximation is no longer valid. In this case, one
can divide the $x$-axis into five regions: (I) $x>x_{1}$, (II)
$x\approx x_{1}$, (III) $x_{1}>x>x_{2}$, (IV) $x\approx x_{2}$,
and (V) $x<x_{2}$. For regions (I), (III), and (V), we still use
the standard WKB wavefunctions,
\begin{eqnarray}
\psi_{I}(x)&=&A_{+}W_{+}(x) +A_{-}W_{-}(x),\nonumber\\
\psi_{III}(x)&=&B_{+}W_{+}(x) +B_{-}W_{-}(x),\nonumber\\
\psi_{V}(x)&=&CW_{+}(x).
\end{eqnarray}
For region (II), we use the linear approximation \cite{bender},
\begin{equation}
V(x)\approx V(x_{1})+V'(x_{1})(x-x_{1}),
\end{equation}
and
\begin{equation}
E^{2}-V(x)\approx \mu_{1}^{2/3}z_{1},
\end{equation}
where
\begin{equation}
\mu_{1}=-V'(x_{1}),\ \ \ \ \ {\rm and}\ \ \ \ \
z_{1}=\mu_{1}^{1/3}(x-x_{1}).
\end{equation}
Then the Schr\"odinger equation becomes
\begin{equation}
\frac{d^{2}\psi}{dz_{1}^{2}}+z_{1}\psi=0,
\end{equation}
with the general solution
\begin{equation}
\psi_{II}(z_{1})=\alpha{\rm Ai}(-z_{1})+\beta{\rm Bi}(-z_{1}),
\end{equation}
where Ai($t$) and Bi($t$) are Airy functions. One can consider
region (IV) in a similar way to obtain
\begin{equation}
\psi_{IV}(z_{2})=\gamma{\rm Ai}(z_{2})+\delta{\rm Bi}(z_{2}),
\end{equation}
where
\begin{equation}
\mu_{2}=-V'(x_{2}),\ \ \ \ \ {\rm and}\ \ \ \ \
z_{2}=\mu_{2}^{1/3}(x-x_{2}).
\end{equation}
Again matching the boundary conditions at $x=\pm\infty$ and the
wavefunctions across $x=x_{1}$ and $x_{2}$, one can obtain the
constants $A_{+}$, $A_{-}$, $B_{+}$, $B_{-}$, $C$, $\alpha$,
$\beta$, $\gamma$, and $\delta$. The transmission probability is
then given by
\begin{equation}
{\cal
T}=e^{-2\int^{x_{1}}_{x_{2}}dx'\sqrt{V(x')-E^{2}}}.\label{e2llvm}
\end{equation}
This is our WKB approximation for the cases with $E^{2}\ll V_{m}$.

\section{Transmission Probabilities for the Dirac Field}

Using the WKB approximations outlined above we shall calculate in
this section the transmission probabilities for the Dirac field in
the radial equation (Eq.~(\ref{V1})) with the potential as given
by Eq.~(\ref{massiveV}).

\subsection{Massless cases}

\begin{figure}[!]
\includegraphics{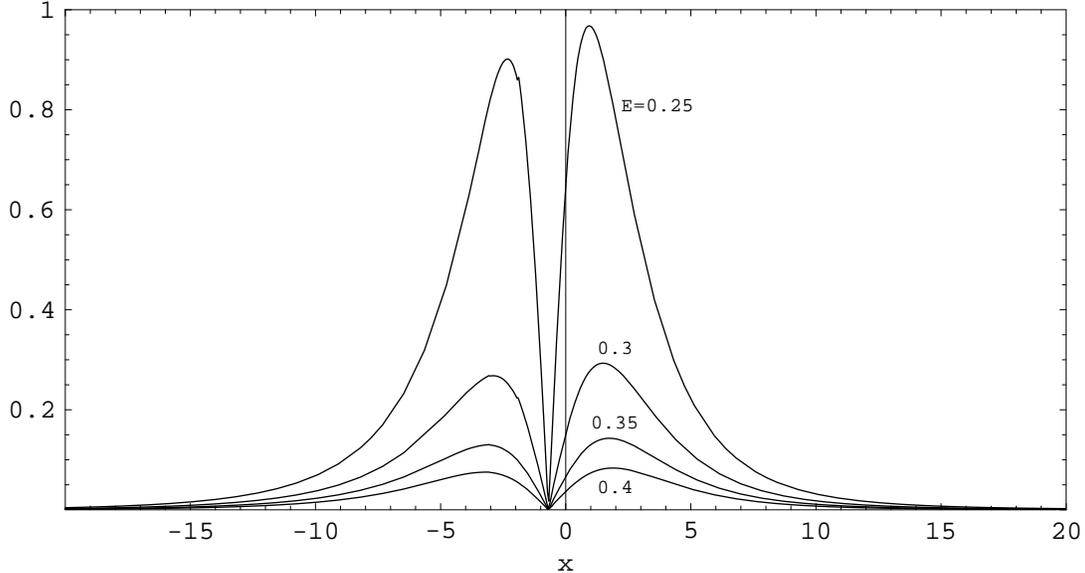}
\caption{\label{validityfig} Validity condition for various values
of $E$ of the massless Dirac fields with $\kappa=1$.}
\end{figure}

\begin{figure}[!]
\includegraphics{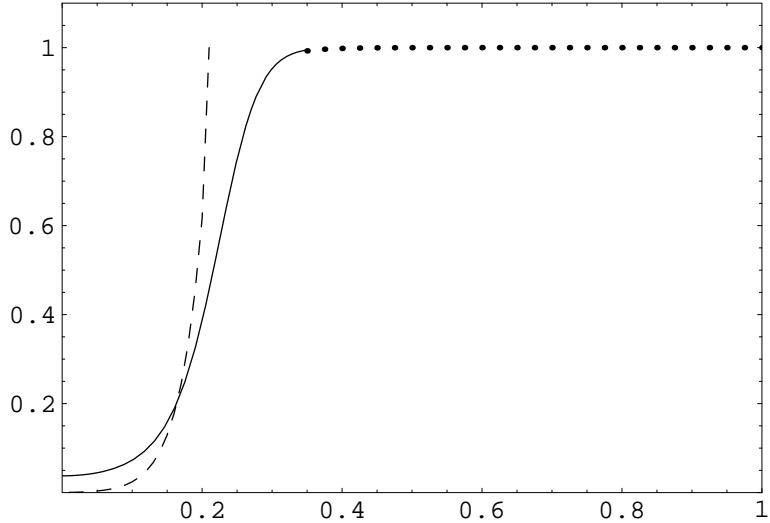}
\caption{\label{kappa1} Transmission probabilities of the massless
Dirac field with $\kappa=1$ in the various WKB approximations for
$E^{2}\gg V_{m}$ (dotted line), $E^{2}\approx V_{m}$ (solid line),
and $E^{2}\ll V_{m}$ (dashed line).}
\end{figure}

First, we consider the massless cases, with the potential
\begin{equation}
V(r,\kappa)=\frac{|\kappa|\Delta^{1/2}}{r^{4}}[|\kappa|\Delta^{1/2}-(r-3)],
\label{m0V}
\end{equation}
where we have used the mass $M$ of the black hole as a unit of
mass and length to simplify the notation so that $\Delta=r(r-2)$.
To proceed we first consider the case with $\kappa=1$. Then we
have
\begin{equation}
\sqrt{V_{m}}=0.216.
\end{equation}
Hence, for $E\gg 0.216$, we can use the formula in
Eq.~(\ref{e2ggvm}) to evaluate the transmission probability. To
check the validity condition, we plot $\left\vert
V'/2p^{3}\right\vert$ for various energy $E$ in
Fig.~\ref{validityfig}. From it, we see that for $E=0.35$, the
maximum value of this quantity,
\begin{equation}
\left\vert\frac{V'}{2p^{3}}\right\vert_{max}\approx 0.1.
\end{equation}
The formula in Eq.~(\ref{e2ggvm}) can therefore be a good
approximation for energy $E$ larger than around 0.35. The
transmission probabilities ${\cal T}$ for $0.35\leq E\leq 1$
calculated from this formula is plotted in Fig.~\ref{kappa1}.

For $E$ smaller than 0.35, one can use the parabolic approximation
for the transmission probability as given in Eq.~(\ref{e2appvm}).
The transmission probabilities for $0\leq E\leq 0.35$ calculated
from this approximation is also plotted in Fig.~\ref{kappa1}. Near
the base of the barrier, the parabolic approximation is no longer
valid because the turning points are far apart. This is apparent
from the fact that the curve tends to around $0.0377$ instead of
zero as $E$ goes to zero.

When the turning points are isolated, one can use the
approximation given in Eq.~(\ref{e2llvm}) for the transmission
probability ${\cal T}$. The transmission probabilities for $0\leq
E\leq 0.216$ calculated from Eq.~(\ref{e2appvm}) are also plotted
in Fig.~\ref{kappa1}. When $E$ is close to $\sqrt{V_{m}}=0.216$,
the formula in Eq.~(\ref{e2appvm}) cannot be trusted as we can see
that ${\cal T}\rightarrow 1$ as $E\rightarrow V_{m}$ in this
approximation.

\begin{figure}[!]
\includegraphics{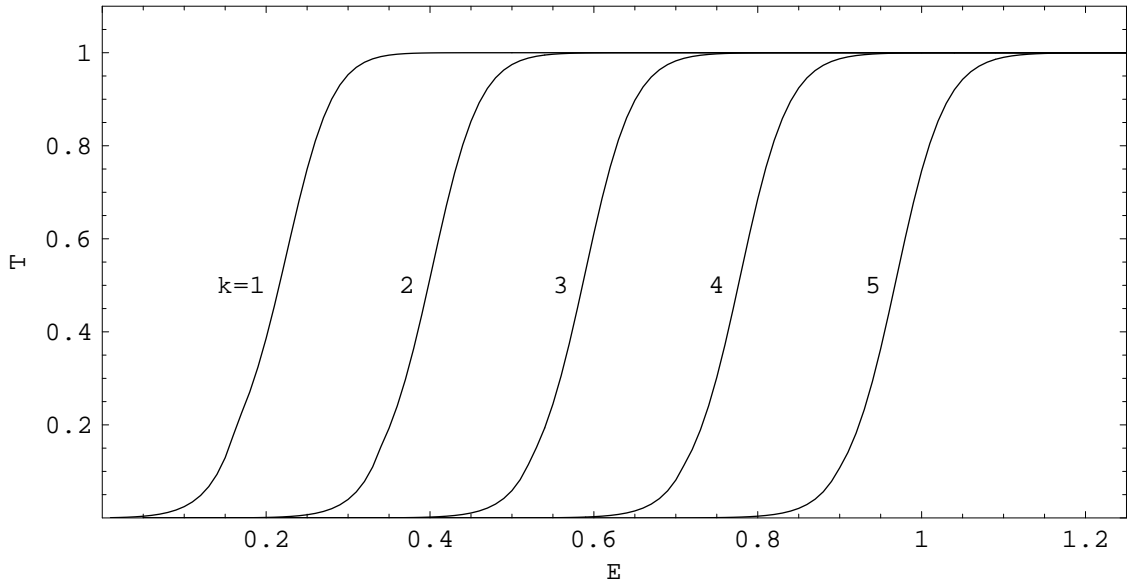}
\caption{\label{kappa1to5} Transmission probabilities ${\cal T}$
of the massless Dirac field with $\kappa=1$ to $5$.}
\end{figure}

In Fig.~\ref{kappa1}, we see that the solid and dashed lines
calculated from these two approximations overlap at around
$E=0.154$. Thus for $E>0.154$, we should take the result with the
parabolic approximation (solid line) and for $E<0.154$, we should
take that of the tunneling approximation (dashed line) instead.
While in the region with $E\approx 0.154$, we can extrapolate
smoothly between these two curves. Combining with the result for
$E>0.35$, we can obtain a curve for the transmission probabilities
in the entire region, $0<E<1$. This is plotted in
Fig.~\ref{kappa1to5}.

In addition to the curve for $\kappa=1$, we also plot the
transmission probabilities for $\kappa=2$, 3, 4, and 5 in
Fig.~\ref{kappa1to5}. These curves are obtained in the same
procedure as outlined above for $\kappa=1$. They are similar to
each other and shift to the right as $\kappa$ increases. The
energy at which ${\cal T}=1/2$ occurs is $\sqrt{V_{m}}$ which
increases as the peak of the barrier gets higher and higher when
$\kappa$ is increased.

\subsection{Massive cases}

\begin{figure}[!]
\includegraphics{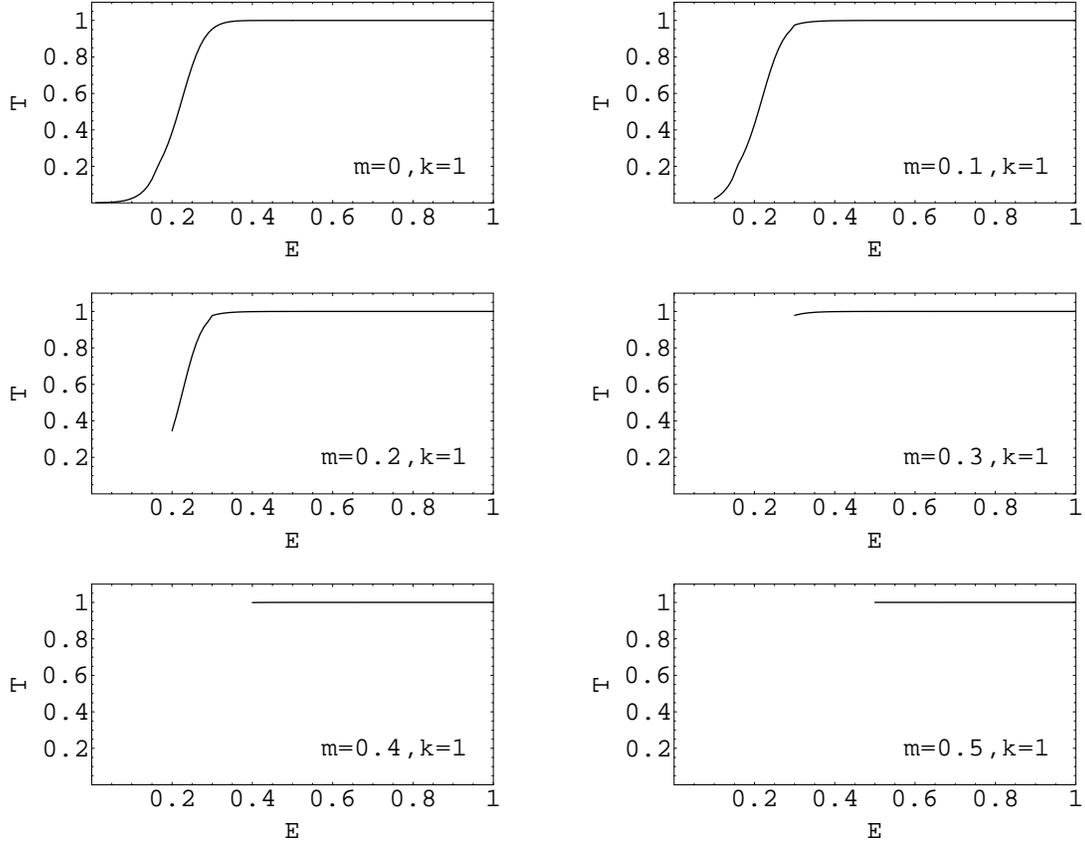}
\caption{\label{massive} Transmission probabilities ${\cal T}$ of
the Dirac field with $\kappa=1$ and $m=0$ to $0.5$.}
\end{figure}

The situations with nonzero $m$ are more complicated because the
potentials can change from a barrier to a step or vice versa as
the parameters $m$ and $\kappa$ are varied as shown in
Figs.~\ref{potentialwithm} and \ref{potentialwithk}. With
$\kappa=1$, the potentials are barriers for $m=0$, 0.1, and 0.2.
One can use the relevant approximations for different values of
the energy $E$, along the same lines as in the massless cases in
the last subsection. The results are plotted in
Fig.~\ref{massive}. However, for $m=0.3$, 0.4, and 0.5, the
potentials are steps. The only approximation one needs is for
$E\gg m$. We also check the validity condition and the result
indicates that the approximation is useful all the way down to
energy value very close to $m$. The results for these masses are
also plotted in Fig.~\ref{massive}. For the potential steps, the
transmission probabilities are almost equal to 1, which are
consistent with the classical results.

\begin{figure}[!]
\includegraphics{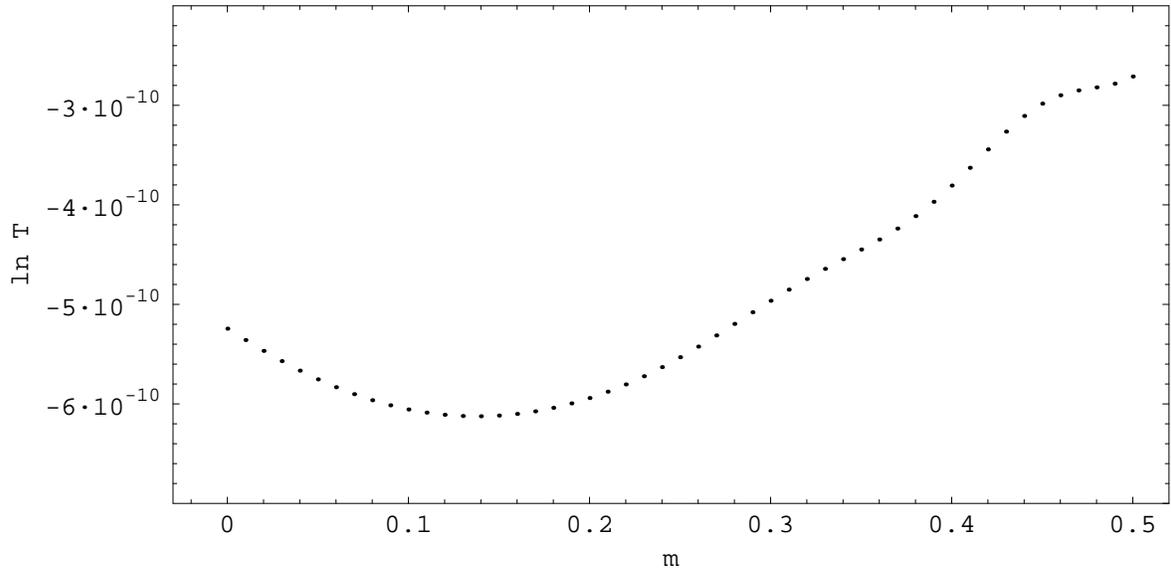}
\caption{\label{logTvsm} Variation of the logarithmic of the
transmission probabilities of the Dirac field ($E=1$ and
$\kappa=1$) with $m$.}
\end{figure}

We also see from the various diagrams in Fig.~\ref{massive} that
the variations of the transmission probabilities ${\cal T}$ with
$m$ are numerically very small. In order to see the changes in
more details, we plotted ${\cal T}$ versus $m$ in
Fig.~\ref{logTvsm} for $E=1$ and $\kappa=1$. Since the
transmission probabilities in these cases are very close to 1, we
take the logarithmic of ${\cal T}$ in the plot. From this figure
we see that ${\cal T}$ first decreases from $m=0$, attends a
minimum around $m=0.14$, and then increases as $m$ is further
increased.

The variations of ${\cal T}$ between $m=0.3$ to $0.5$ are quite
unexpected. This is because with these values of $m$, the
potentials are in the form of steps with larger and larger step
height when $m$ is increased. For simple potential steps, it is
known that the transmission probabilities decrease as the steps
get higher and higher for fixed energy. We thus see that for the
black hole effective potentials, the transmission probabilities
are not determined only by the height of the potential but also
the overall shape of it.

\begin{figure}[!]
\includegraphics{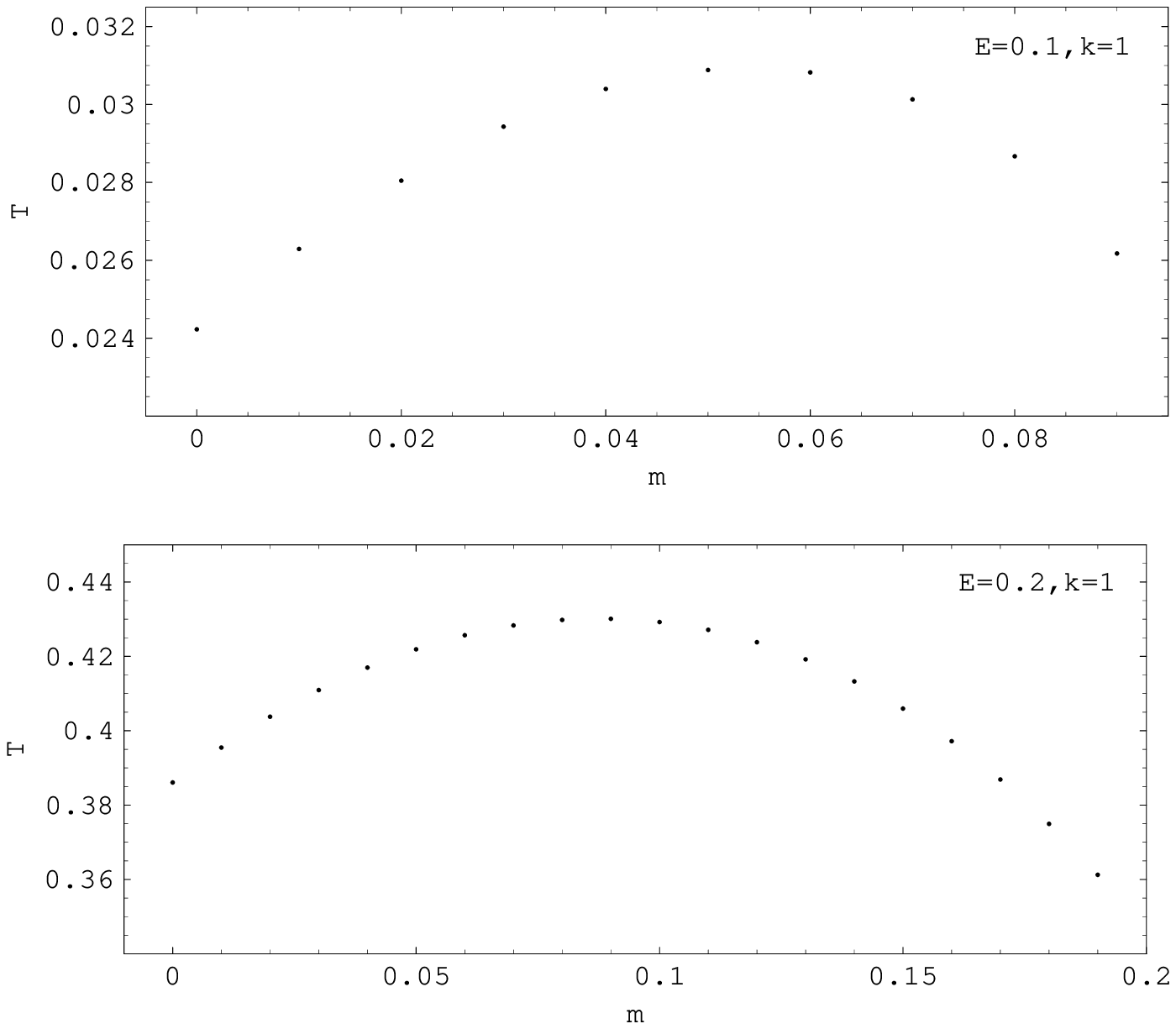}
\caption{\label{TvsmsmallE} Variation of the transmission
probabilities of the Dirac field with $m$ for $E=0.1$ and $0.2$.}
\end{figure}

In Fig.~\ref{logTvsm}, $E=1$ and is much larger than
$\sqrt{V_{m}}$. For energy values closer to $\sqrt{V_{m}}$, the
variations can be quite different. As shown in
Fig.~\ref{TvsmsmallE}, for $E=0.2$, that is, when $E^{2}$ is near
the peak of the potential, ${\cal T}$ increases first as $m$ is
increased from $0$, attends a maximum value around $m=0.09$, and
then decreases. The same trend is found for $E=0.1$ where it is
well below the peak of the potential. However, the maximum value
in this case is at around $m=0.05$.

As we can see from the above discussions, the transmission
probability ${\cal T}$ in general increases with the mass $m$ when
$m$ is large enough that the effective potential is in the form of
a step. However, when $m$ is smaller and the effective potential
is in the form of a barrier, the variations of ${\cal T}$ with $m$
can be quite complicated. When $E$ is large and well above the
peak, ${\cal T}$ first decreases and then increases when $m$ is
increased from $m=0$. When $E$ is small with its value near or
well below the peak of the potential, the variation is reversed,
that is, ${\cal T}$ first increases and then decreases when $m$ is
increased from $m=0$.

\section{Conclusions and Discussions}

We study the radial equations of the massive Dirac field in the
spherically symmetric Schwarzschild spacetime. Using the WKB
approximations and the appropriate connection formulae, we are
able to give semianalytic formulae for the transmission
probability ${\cal T}$ of the radial wavefunction with the energy
$E^{2}\gg V_{m}$, $E^{2}\approx V_{m}$, and $E^{2}\ll V_{m}$,
$V_{m}$ is the maximum value of the effective potential. For the
massless cases, we find that, as shown in Fig.~\ref{kappa1to5},
the variations of ${\cal T}$ with the energy $E$ and the angular
momentum number $\kappa$ are similar to that for the scalar case
\cite{futterman,sanchez}. Since the potentials are in the form of
barriers and the heights of the peaks of these barriers increase
with $\kappa$, ${\cal T}$ for fixed $E$ will thus decrease as
expected. This means that waves with lower angular momenta but
with fixed energy will be absorbed more easily by the black hole.

The massive cases are more complicated. The effective potentials
in these cases change from barriers to steps when the mass of the
field is increased. When the potential is still in the form of a
step and the energy $E$ of the field is well above the maximum
value of the potential, the transmission probability ${\cal T}$
decreases with $m$. Thus higher mass fields will get absorbed by
the black hole more easily. However, when $m$ is decreased to such
an extend that the potentials become barriers, the variation trend
changes as shown in Fig.~\ref{logTvsm}. At some value of $m$
(around 0.14 for $E=1$ and $\kappa=1$), ${\cal T}$ turns around
and increases when $m$ is further decreased to $0$. Therefore, we
see that the variations of ${\cal T}$ with $m$ is complicated when
the potentials are in the form of barriers. This is also true when
$E$ is smaller with its value near or well below the peak of the
potential as shown in Fig.~\ref{TvsmsmallE}.

After calculating the transmission probabilities for the radial
wavefunctions, one can evaluate the corresponding phase shifts and
cross sections in various scattering situations \cite{futterman}.
In the semiclassical limit one can also study the interesting
phenomena such as black hole glories \cite{matzner}, orbiting and
spiraling scatterings \cite{anninos} by deriving the semiclassical
deflection function \cite{ford}. We hope to further investigate
these issues in our future publications.

Quite peculiarly for black holes, one can also study the
absorption cross sections for various wave fields. There has been
quite a lot of interest in these cross sections in relation to the
higher-dimensional black holes in string theory, especially the
low-energy absorption cross sections \cite{das}. However, the WKB
approximations that we use in this work is not adequate at low
energy, that is, when $E\approx m$. One can nevertheless improve
the WKB approximations in these threshold situations when $E$ is
near the top of the step or when it is near the base of the
barrier \cite{eltschka,moritz}. This improved approximations may
therefore provide an alternative to the usual method \cite{unruh}
in obtaining these low-energy absorption cross sections.

\begin{acknowledgments}
This work is supported by the National Science Council of the
Republic of China under contract number NSC 91-2112-M-032-011.
\end{acknowledgments}

\bibliography{DiracWKB}

\end{document}